\documentclass[twocolumn,aps,prl,floatfix,superscriptaddress,nobibnotes,nofootinbib,altaffilsymbol]{revtex4-2}
\usepackage{amsmath,amssymb}
\usepackage{graphicx,float}
\usepackage{times,txfonts}
\usepackage{color}
\usepackage{multirow}
\usepackage{bbold}
\usepackage{color}
\usepackage{soul}
\usepackage{hyperref}
\usepackage{times,txfonts}
\usepackage{natbib}
\usepackage{nicefrac}
\usepackage{blindtext}
\usepackage[export]{adjustbox}
\usepackage{mathrsfs}
\usepackage{textcomp}
\usepackage{blkarray}
\usepackage{multirow}
\usepackage[normalem]{ulem}
\usepackage{tabu}

\usepackage{hyperref}
\hypersetup{
   pdfpagemode=None,
   pdfstartpage=1,
   pdfmenubar=true,
   pdftoolbar=true,
   colorlinks = true,
   linkcolor=blue,
   citecolor=blue,
   urlcolor=blue,
   bookmarksopen=false
 } 

 \usepackage{booktabs}

\definecolor{lgreen}{RGB}{15,150,15}

\newcommand{\tj}[1]{{\color{magenta}#1}}

\begin{document}

\title{Full-Field Mode Sorter for Optical Knots}

\author{Tareq Jaouni}
\affiliation{Nexus for Quantum Technologies, University of Ottawa, K1N 5N6, Ottawa, ON, Canada}
\email{tjaouni@uottawa.ca}

\author{Roohollah Ghobadi}
\affiliation{Nexus for Quantum Technologies, University of Ottawa, K1N 5N6, Ottawa, ON, Canada}

\author{Ebrahim Karimi}
\affiliation{Nexus for Quantum Technologies, University of Ottawa, K1N 5N6, Ottawa, ON, Canada}
\affiliation{Institute for Quantum Studies, Chapman University, Orange, California 92866, USA}

\begin{abstract}
Optical knots are topologically structured light fields whose phase or polarization singularities trace linked or knotted trajectories during propagation, making them promising candidates for high-dimensional optical information carriers. Their use in communication or quantum-information protocols, however, requires a practical readout method that can distinguish a chosen knot alphabet with low crosstalk. Here, we demonstrate a proof-of-principle full-field sorter for optical knots using one or two optimized phase-only elements. The sorter maps each input knot to a predefined output region and is optimized directly from the output intensity distributions to enhance correct assignment, suppress crosstalk, and avoid degenerate mappings between distinct knots. We apply the method to an alphabet composed of the Hopf link, trefoil, and cinquefoil optical knots. Two optimized phase planes improve the sorting performance relative to a single plane and enable high distinguishability for the three-knot alphabet. We further benchmark the sorter under common experimental imperfections. These results extend full-field optical mode sorting to topologically structured light and provide a readout route for knot-based high-dimensional optical communication.
\end{abstract}

\maketitle
\section{Introduction}

High-dimensional photonic quantum information remains highly sought after, promising increased information capacity and noise tolerance, which altogether enhances its practicality in real-world settings~\cite{bechmann-pasquinucci_quantum_2000}. However, choosing the best photonic degree of freedom remains context-dependent: their robustness to spurious channel fluctuations, ease of experimental implementation and distinguishability all determine the choice of encoding alphabet. Structured light has emerged as a strong candidate for enabling high-dimensional optical alphabets for quantum key distribution, with orbital angular momentum (OAM) providing a widely used example ~\cite{bliokh2023roadmap,sit_high-dimensional_2017, bouchard_quantum_2018, bouchard2018experimental, mirhosseini2015high, jaouni_predicting_2025, willner2015optical}. Their use in information tasks requires some method to measure the optical alphabet encoded in spatial modes. To this end, scientists have employed optical mode sorters. Existing sorters have separated spatial modes using interferometric transformations~\cite{gu_gouy_2018, leach_measuring_2002}, scattering systems~\cite{fickler2017custom}, and multi-plane or full-field diffractive architectures~\cite{fickler_paper, fontaine_laguerre-gaussian_2019, a_rocha_self-configuring_2025, kupianskyi_high-dimensional_2023, lib2025high}.

Nontrivial three-dimensional topologies, such as Möbius strips, ribbon strips, and knots, are demonstrated to be generated in the optical domain~\cite{bauer2015observation,bauer2019multi,leach_vortex_2005,larocque2018reconstructing,larocque2020optical}. Optical knots offer a topologically structured candidate, in which phase singularities trace closed loops in three-dimensional space~\cite{berry_knotted_2001,sugic2020knotted,ferrer-garcia_polychromatic_2021, leach_vortex_2005, berry_knotting_2001}. The topology can be traced to a corresponding braid representation, motivating optical knots as carriers of topological information ~\cite{larocque2020optical,ferrer2022secure,freedman2003topological}. Furthermore, their topologically structured field configurations have motivated studies on their robustness to experimental defects and turbulent channels~\cite{ak_paper, pires_stability_2025}.

Devising a measurement for optical knots is not readily apparent, even when one considers only their transverse field profile. Unlike other spatial modes, optical knots lack a clear modal index that mode-specific analytic sorters can exploit. Moreover, they are generally non-orthogonal optical fields, which constrains the effectiveness of any given optical mode sorter~\cite{helstrom_quantum_1969}. One or two optimised phase transformations are capable of sorting general transverse modes into prescribed output channels, without relying on a mode-specific analytic sorter~\cite{fickler_paper}. This full-field strategy is, therefore, well suited to optical knots, whose information is not contained in a single modal index but in a complex field profile and its propagation-dependent singular structure.

Here, we demonstrate a full-field sorter for optical knots by adapting the optimized phase-plane architecture of Ref.~\cite{fickler_paper}. The device uses one or two phase-only elements to sort distinct optical knots to predefined output regions. We optimize the phase masks directly from the output intensity distributions, balancing correct assignment, crosstalk suppression, worst-channel performance, and distinguishability of the output probability matrix. We apply the sorter to an alphabet of Hopf link, trefoil, and cinquefoil optical knots. We compare one- and two-plane architectures and test the optimized sorter against rotations, transverse translations, phase aberrations, and experimentally motivated channel distortions.

\begin{figure*}
\includegraphics[width=\textwidth]{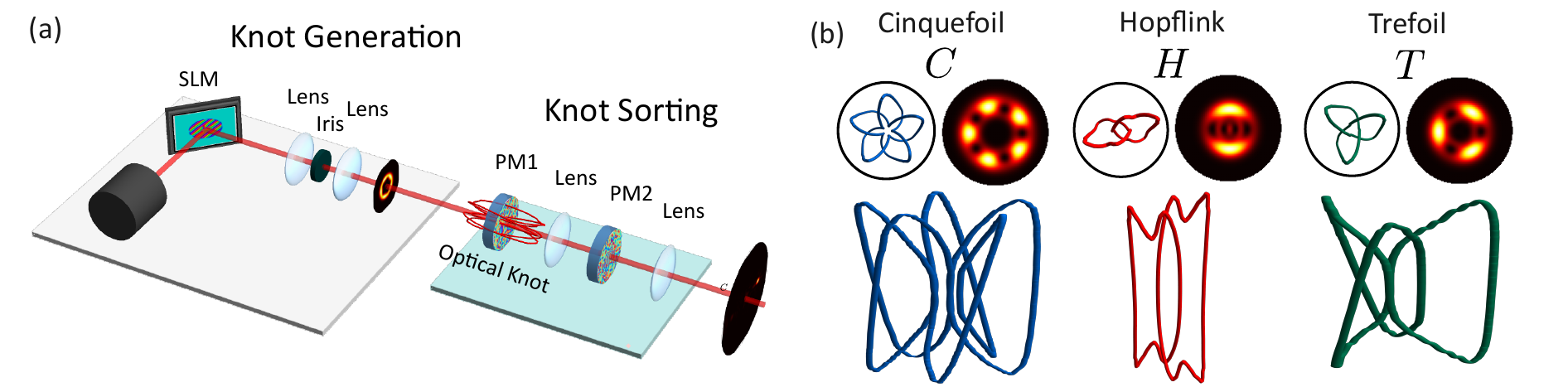}
    \caption{\textbf{Experimental concept.} (a) We first generate the desired knotted field by encoding the required phase pattern onto an SLM and selecting for the first diffraction order using a 4f imaging system. We then apply the optimized phase masks onto the field at the waist transverse plane. (b) The knot alphabet is used to verify the sorting performance. We show the knotted fields' intensity profile at its waist plane -- $|\psi(\rho, \phi, z=0)|^{2}$ -- as well as each knot's corresponding singular skeleton. All optical knots have the same Gaussian waist envelop parameter of $s=1.0$. The shape parameters for each optical knot are $(a,b) = (2.0, 2.0)$ for cinquefoil, $(0.9, 0.9)$ for trefoil, and $(0.6, 0.6)$ for Hopf link. }
    \label{fig:expt_sorting}
\end{figure*}

\section{Theory}

\textit{Optical-knot alphabet:} We consider an alphabet of $d$ paraxial optical-knot fields defined at a reference transverse plane $z=0$. Each field is represented by a complex scalar envelope $\psi_n(\mathbf r)\equiv \psi_n(x,y)$, with $n=1,\dots,d$, and $\mathbf r=(x,y)$. The relevant inner product is taken over the finite numerical or experimental aperture $\mathcal A$:
$\langle f|g\rangle_{\mathcal A}=\int_{\mathcal A} f^*(\mathbf r)g(\mathbf r)d^2\mathbf r$. The knot fields are normalized as $\langle \psi_n|\psi_n\rangle_{\mathcal A}=1$. At the waist plane, the fields are generated from Gaussian-weighted Milnor polynomials,
\begin{equation}\label{psin}
u(\rho,\phi)=\mathcal N
\exp\left[-\frac{\rho^2}{2s^2}\right]
P(\rho,\phi;a,b),
\end{equation}
where \(P(\rho,\phi;a,b)=\sum_{l} c_{l}(\rho,a,b) e^{il\phi}\) is the polynomial defining the chosen knot family, \(s\) sets the Gaussian envelope width, \(a,b\) control the knot geometry, and \(\mathcal N_n\) is a normalization constant~\cite{leach_vortex_2005}. The explicit polynomial forms for the trefoil, cinquefoil, and Hopf link are listed in the Appendix.

\textit{Phase-only sorter as a linear optical measurement:} The sorter consists of one or two phase-only modulation planes separated by propagation. A phase mask acts as the multiplicative operator $\Phi_j(\mathbf r)=\exp{(i\varphi_j(\mathbf r))}$ where \(\varphi_j(\mathbf r)\) is the optimized phase pattern on the \(j\)-th plane. For a single phase plane followed by a lens, the field transformation is modelled as $U_1=\mathcal F [\Phi_1]$, where \(\mathcal F\) denotes Fourier propagation to the back focal plane. For two phase planes, the transformation is $U_2=\mathcal F [\Phi_2 \mathcal F [\Phi_1]]$. More compactly, we write the sorter transformation as \(U\), with \(U=U_1\) or \(U=U_2\) depending on the implemented architecture. For an input knot \(\psi_n\), the output field is $\Psi_n^{\mathrm{out}}(\mathbf r)=(U\psi_n)(\mathbf r)$. Each input knot is assigned to a target detection region \(\mathcal {W} _ {n} \) in the output plane. The detected intensity in output channel \(m\) for input \(n\) is
\begin{equation}\label{Imn}
I_{m n}=\int_{\mathcal W_m}
\left|
\Psi_n^{\mathrm{out}}(\mathbf r)
\right|^2
d^2\mathbf r.
\end{equation}

Equivalently, introducing the projector $\Pi_m=\int_{\mathcal W_m}|\mathbf r\rangle\langle \mathbf r|d^2\mathbf r$, we may write $I_{mn}=\langle \psi_n|U^\dagger \Pi_m U|\psi_n\rangle$. Thus, the sorter implements a spatial measurement with effects $E_m=U^\dagger \Pi_m U$. This formulation makes clear that the device is not merely redirecting beams; it realizes a finite-outcome optical measurement on the input field alphabet.

\begin{figure*}
\includegraphics[width=\textwidth]{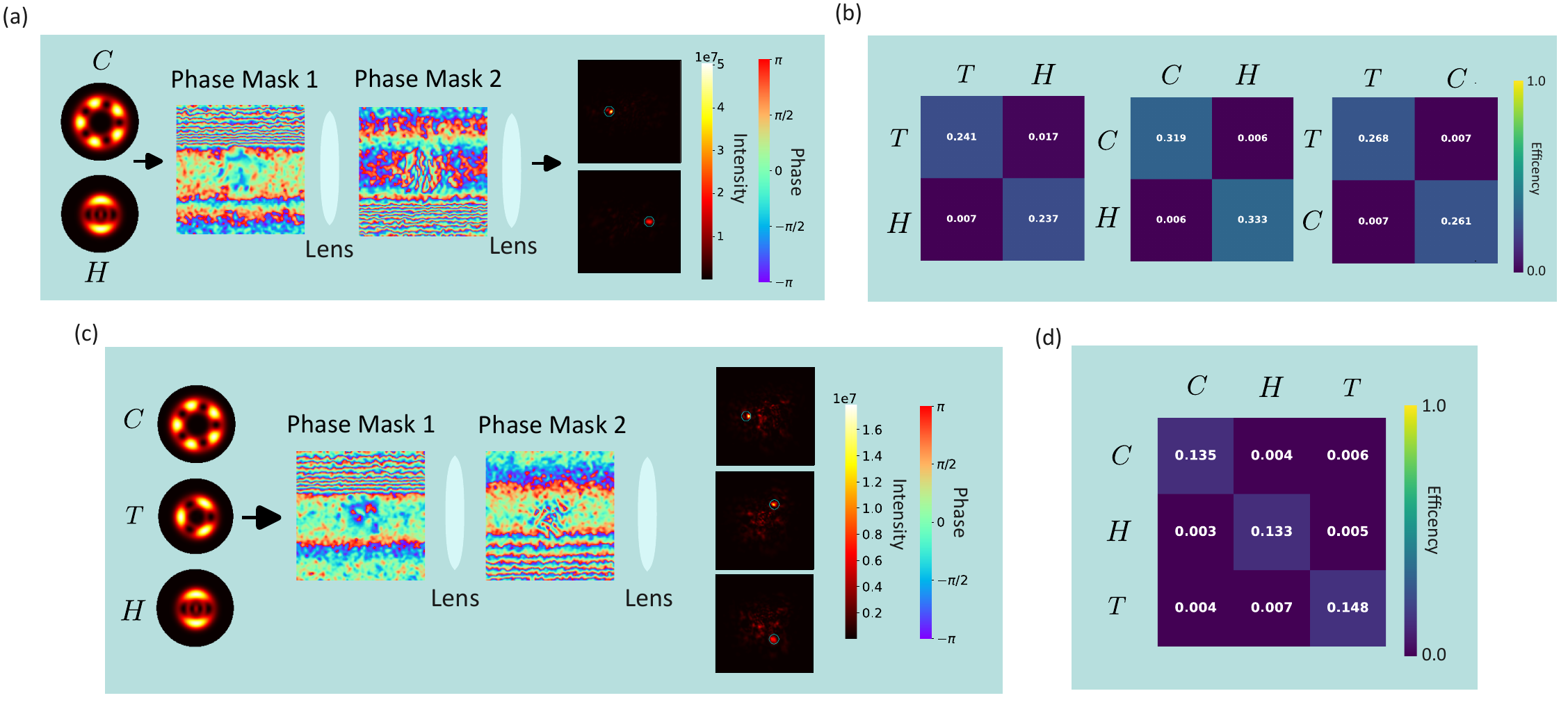}
    \caption{\textbf{Results for sorting knots using two optimized phase elements.} In (a)-(b), we demonstrate our sorter's ability to sort $d=2$ knotted optical modes. (a) Optimized phase elements and output field for sorting cinquefoil (C) and Hopf link (H) optical knots. (b) Crosstalk matrices for all possible two-knot combinations in our alphabet. In (c)-(d), we apply our sorting routine on d=3 knots, utilizing our entire knot alphabet. (c) optimized phase elements and output field for sorting the knots (d) crosstalk matrix associated with our sorter. }
    \label{fig:two_plane}
\end{figure*}

\textit{Optimization objective:}  A useful knot sorter must satisfy several distinct requirements. It must direct each input knot to its assigned output channel, suppress leakage into the wrong channels, perform reliably for every knot in the alphabet, and avoid degenerate mappings in which different knots produce the same output pattern. Since the phase masks are optimized with a genetic algorithm, these requirements must be compressed into a single scalar fitness function. We construct this objective from three factors: a balanced sorting contrast $C_b(\alpha)$, a communication-motivated error penalty $R_d(e_b)$, and a determinant-based distinguishability factor $V_p$. All three are obtained from Eq.(\ref{Imn}). 

We first convert the intensity matrix $I_{mn}$ into a conditional assignment matrix, 
\begin{equation}
p_{m|n}=\frac{I_{mn}}
{\sum_{m=1}^d I_{mn}},
\end{equation}
with $\sum_{m}p_{m|n}=1$ for each input $n$. Thus $p_{m|n}$ is the probability that input knot $n$ is assigned to output channel $m$, conditioned on detection inside one of the accepted output windows.

For each input knot, we define the contrast as, $$C_n=p_{n|n}-\frac{1}{d-1}\sum_{m\neq n} p_{m|n},$$ where the first term rewards assignment to the correct channel, while the second term penalizes leakage into all incorrect channels. Thus, $C_n=1$ for ideal sorting and $C_n=0$ for a uniform random assignment. To ensure that the sorter works across the entire alphabet, not only for the easiest knots, we combine the mean contrast with the worst-channel contrast,
\begin{equation}
C_{\text{b}}(\alpha)=\alpha(\text{min}_{n} C_n)+(1-\alpha)\bar{C},
\end{equation}
where the mean contrast $\bar{C}=\frac{1}{d}\sum_{n}C_n$ quantifies the overall sorting quality, while the bottleneck contrast $\text{min}_{n} C_n$ identifies the weakest sorted knot. The parameter $0<\alpha\leq1$ controls the relative weight of this worst-channel penalty.

If the knots are used as alphabet symbols, assigning input $n$ to any output channel $m\neq n$ corresponds to decoding the wrong symbol. The average symbol error rate is $e_b=\frac{1}{d}\sum_{n=1}\sum_{m\neq n}^d p_{m|n}$. Equivalently, in term of average sorting probability $p_{\mathrm{sort}}=\frac{1}{d}\sum_{n=1}^d p_{n|n}$, one has 
$e_b=1-p_{\mathrm{sort}}$. We then define the d-dimensional communication-motivated penalty 
\begin{equation}
    R_d(e_b)=\max\left[0,\log_2 d-2h_d(e_b)\right],
\end{equation}
where $h_d(e_b)=-e_b\log_2\left(\frac{e_b}{d-1}\right)-(1-e_b)\log_2(1-e_b)$. This quantity represents the secret-key rate for high-dimensional states.  

To penalize solutions in which multiple input knots are mapped onto the same output channel, we introduce a determinant-based distinguishability factor $V_P=|\det p|$. For a perfectly sorted alphabet, $p$ is the identity matrix and \(V_p=1\). If two inputs produce identical conditional output distributions, two columns of \(p\) become linearly dependent and \(V_p=0\). Thus, \(V_p\) penalizes the collapse of the alphabet into fewer distinguishable output patterns. The optimization objective is then taken as,
\begin{equation}\label{eq:optim_obj}
    F=C_{b}(\alpha)
    R_d(e_b)V_P^\gamma,
\end{equation}
where \(0\leq \alpha\leq1\) controls the emphasis on the worst-performing channel, and \(0\leq\gamma\leq1\) controls the strength of the determinant penalty. 

The modifications to the metric formulated in~\cite{fickler_paper} are owing to the non-orthogonality of the optical knotted fields. While one is capable of finding phase elements capable of sufficiently sorting knotted fields, the partial confusion of optical knots into incorrect channels, depending on the knots' pairwise overlap, is significantly more probable than with an orthogonal sorting alphabet. This can result in columns of $p$ that are linearly dependent, even if $p$ is not diagonal, leading to $\det{p} \rightarrow 0$ and, ultimately, a harsher penalty on solutions that partially sort the optical knots across incorrect channels. The inclusion of additional penalties enables the retrieval of phase elements that, over fewer iterations, sort optical knots more efficiently and with less modal crosstalk.

\textit{Numerical simulation:} We simulate the experimental procedure, see the sketched in Fig.~\ref{fig:expt_sorting}-(a), which sorts the optical knots. We create the desired knot by encoding the knotted field using a Spatial Light Modulator (SLM)~\cite{leach_vortex_2005,ak_paper}. We superimpose the knotted field with a Gaussian beam and verify the singular structure using phase-shifting digital holography. We modulate the optical knot with the first phase mask and observe the modulated field at the back focal plane of the first lens via fast Fourier transform. All of our simulations are done using a $780$~nm wavelength beam with a beam waist of $100~\mu$m and a $128 \times 128$ numerical window. We define the transverse beam profile and phase elements with a spatial resolution of $20~\mu$m, reflecting the pixel size of commercially available SLMs.

We obtain the optimized phase masks using a Genetic Algorithm (GA). We direct our phase masks to sort incoming knots onto circular spots prearranged in a circle of radius $1.0$ mm. We choose an optimization objective with $\gamma=1.0$ and $\alpha=0.5$, as we find that this configuration achieves higher fitness values over less generations. At each iteration, we also apply a Gaussian filter over the phase masks. Their strength is fine-tuned to minimize scattering while also preserving modulation resolution. 

\section{Results}
\begin{figure*}
\includegraphics[width=\textwidth]{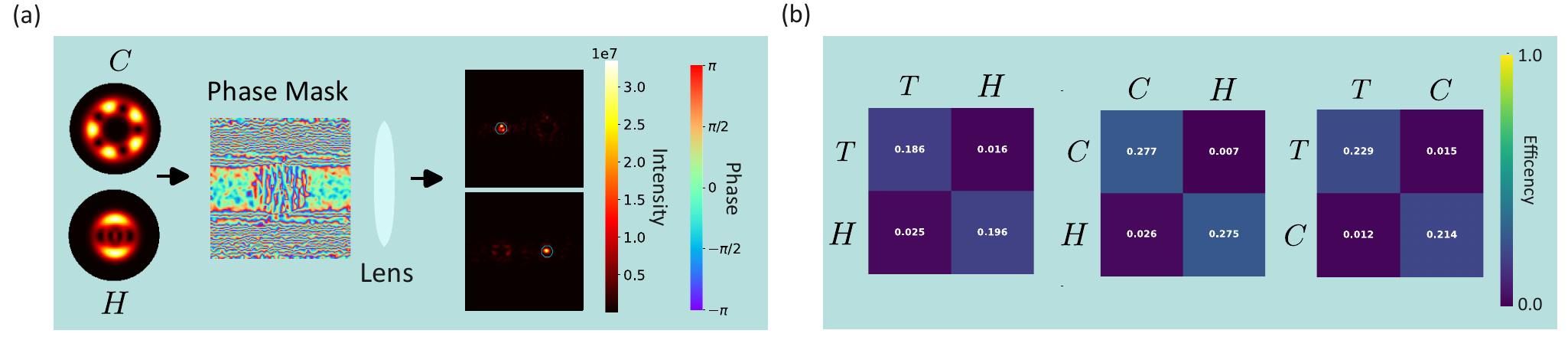}
    \caption{\textbf{Results for sorting knots using one optimized phase plane.} We demonstrate the sorting performance on two-knot combinations. (a) Optimized phase elements and output field for sorting cinquefoil (C) and Hopf link (H) optical knots. (b) Crosstalk matrices for all possible two-knot combinations in our alphabet.}
    \label{fig:one_plane}
\end{figure*}

We attempt to sort our Knot Alphabet with up to two optimized phase elements. To evaluate our optimized sorters' effectiveness, we calculate the channel intensities at the output plane and report each channel's efficiency, $I_{nn}$, for the $n^{th}$  mode. For each channel, we also calculate the normalized sorter probability, $P_n = I_{nn}/({I_{nn} + \sum_{m \neq n}I_{nm}})$. We attribute to each optimized sorter the sorting ability, $P = \bar{P_{n}}$, by computing the mean sorting probability. We first investigate our sorter's ability to sort $d=2$ knot combinations, then, with the entire $d=3$ Knot Alphabet, determine how performance scales with higher dimensionality. All optimized phase elements exhibit trace amounts of scattering due to imperfections in the numerical optimization routine.

\textit{One phase plane:} We start by using an optimized phase element to sort two knots. In Figure~\ref{fig:one_plane}-(a), we show the GA-optimized hologram obtained after $2.0 \times 10^{5}$ iterations to sort the cinquefoil (C) and Hopf link (H) knot, and the resulting field at the image plane. In all cases, the element diverts the incident knotted field to its corresponding channel. For all mode combinations, we report efficiencies of $18\% - 28\%$ and $P > 90\%$. Each hologram's effectiveness is correlated with the overlap between the sorted modes: the hologram for knots C and H, which exhibits the smallest overlap of all the combinations, demonstrates the highest efficiency and sorting ability, whereas knots T and H have the highest modal overlap but the lowest efficiency and sorting ability. The optimized holograms themselves adapt a grating-like phase pattern, which is reminiscent of phase-flattening techniques utilized for OAM modes~\cite{mair2001entanglement,qassim2014limitations,trichili2016optical}. We also find phase planes that sort all optical knots in the Alphabet. We report efficiencies of $7\%$ and sorting ability of $80\%$. This hints at a fundamental limitation attributed to using one optimized phase element in sorting multiple non-orthogonal modes.

\textit{Two-phase plane:} We repeat our experiments using two-phase plane elements. We show results for $d=2$ and $d=3$ mode sorting in Figure~\ref{fig:two_plane} with two phase planes optimized after $1.5e5$ iterations. For $d=2$ knot combinations, we observe efficiencies of $25\% - 35\%$ and sorting abilities of $P > 95\%$. As with using a single-phase plane, we observe overlap-dependent bounds on the performance of each input knot combination. For $d=3$, we report efficiencies of $13\%$ and sorting ability of $93\%$. Though sorting efficiency decreases with increasing modes, sorting ability remains robust. This suggests that we may sort a highly-dimensional Knot Alphabet with minimal cost in sorting ability. 

\subsection{Robustness to input perturbation}

\begin{figure*}
\includegraphics[width=\linewidth]{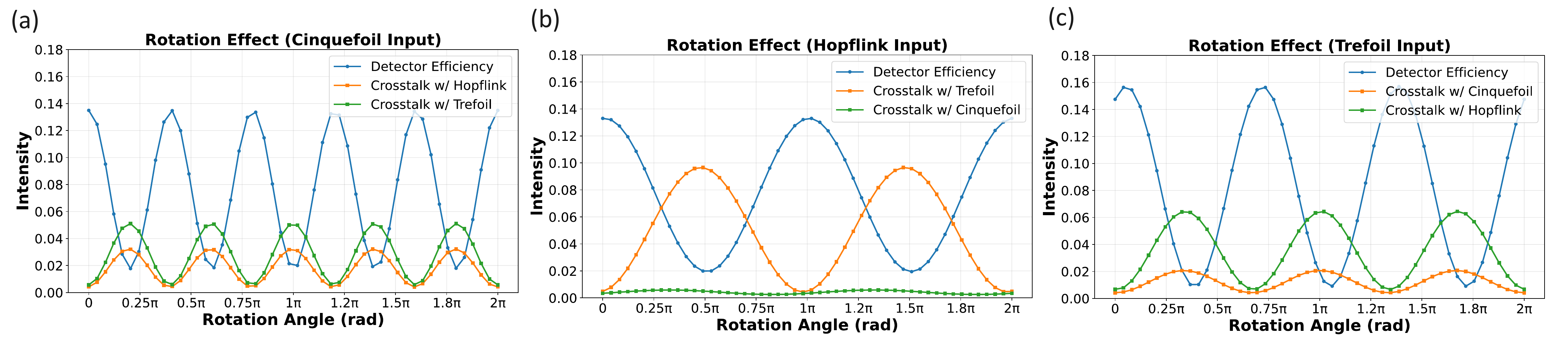}
    \caption{\textbf{Impact of rotating the input field.} We plot the detection efficiency and modal crosstalk as a function of the rotation angle $\theta$. We consider 50 points over the range $\theta \in [0, 2\pi]$. We report how these quantities change when sending (a) cinquefoil (b) Hopf link and (c) trefoil inputs.}
    \label{fig:sort_rotation}
\end{figure*}

After optimizing the phase masks for the ideal knot alphabet, we keep the sorter of Figure~\ref{fig:two_plane}-(b) fixed and perturb only the input field. This allows us to test whether the optimized transformation remains effective under realistic deviations between the incident knot and the field used during training. We consider three classes of perturbations: (1) relative rotations, (2) transverse translations, and (3) phase aberrations. Recomputing the output intensities
\begin{equation}\label{Imnp}
I_{m n}(\lambda)=\int_{\mathcal W_m}\left|U_{*}\Psi_n^{(\lambda)}(\mathbf r)\right|^2 d^2\mathbf r,
\end{equation}
where $U_{*}$ is the fixed sorter, for each perturbation value $\lambda$, and defining a corresponding conditional assignment matrix $p_{m|n}(\lambda)$. \newline
\begin{figure*}
\includegraphics[width=0.90\linewidth]{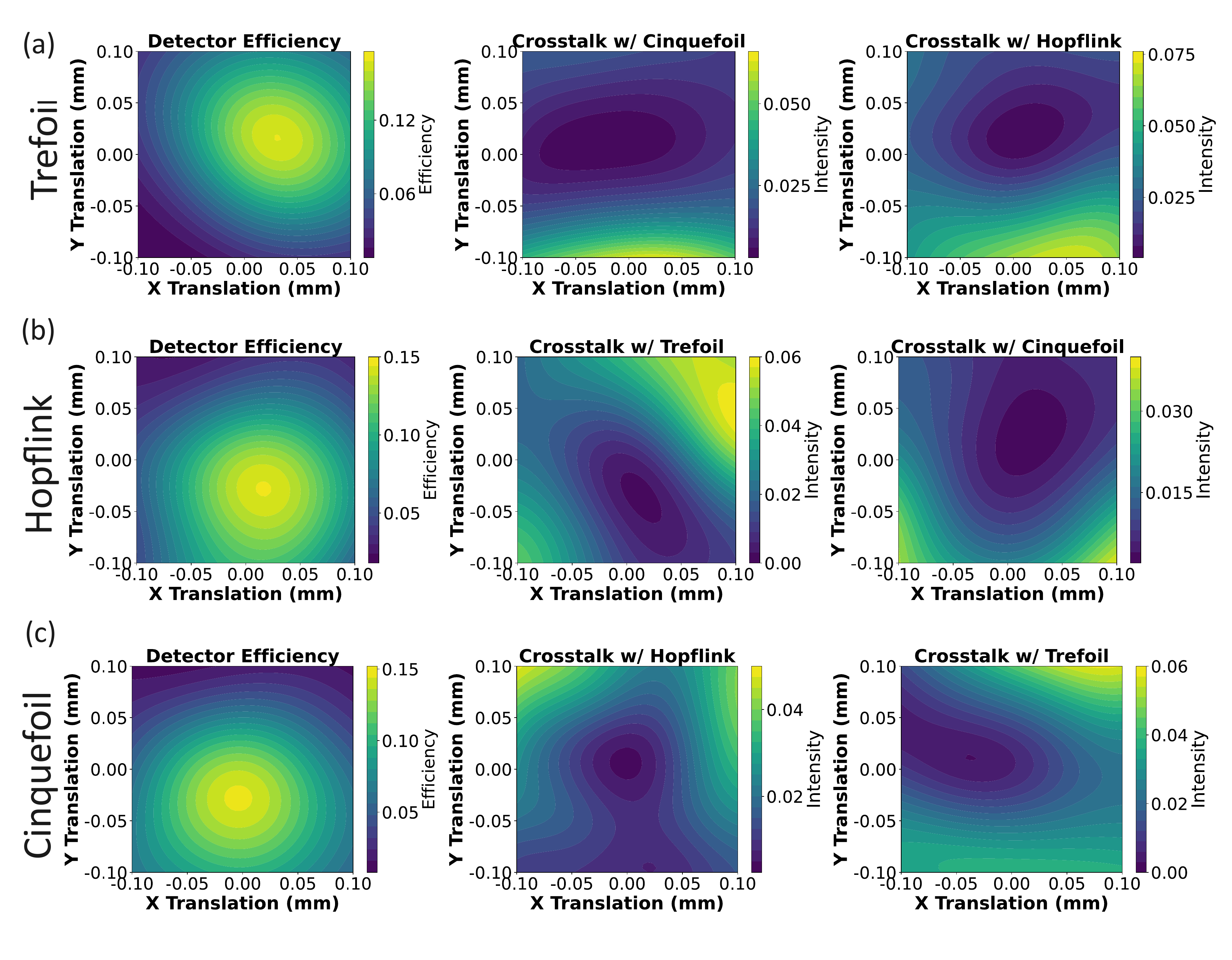}
    \caption{\textbf{Impact of translating the input field on sorting performance.} In (a)-(c), we show heatmaps of the detection efficiency and modal crosstalk as a function of $(x_{trans}, y_{trans})$. We report how these change when sending (a) trefoil, (b) Hopf link, and (c) cinquefoil as inputs to the mode sorter. }
    \label{fig:sort_translation}
\end{figure*}

\textit{(1) Rotation:} A relative rotation between the incoming knot and the optimized phase masks is modeled by rotating the input field before it enters the sorter, $\psi_n^{(\theta)}(x,y)=\psi_n(x',y')$ with $x'=x\cos\theta+y\sin\theta$ and $y'=-x\sin\theta+y\cos\theta$.
 
Figure~\ref{fig:sort_rotation} demonstrates how the overall sorting performance changes depending on the input. Since rotation multiplies each angular sector by $e^{-il\theta}$, $I_{mn}$ contains Fourier components at frequencies $l-l'$, for $l,l'\in\{0,\pm n\}$. Thus, the expected modulation is twofold for the Hopf link, threefold for the trefoil, and fivefold for the cinquefoil. Moreover, the amplitudes of the crosstalk between different knot pairs correspond to their overlap: Figure~\ref{fig:sort_rotation}(b), for example, exhibits a significantly higher crosstalk modulation with the trefoil knot than with the cinquefoil knot, which reflects the differences in the overlaps.

\textit{(2) Translations:} We then apply relative translations to the input knotted beam. This may be of interest when considering small offsets in the input knotted beam relative to the optimized phase elements of the sorter. We translate the input field before it enters the sorter,  $\psi_n^{(trans)}(x,y)=\psi_n(x',y')$ with $x'=x+x_{trans}$, $y'=y+y_{trans}$, and $x_{trans}, y_{trans} \in [-0.1 \text{mm}, 0.1 \text{mm}]$. 
\begin{figure*}
\includegraphics[width=\linewidth]{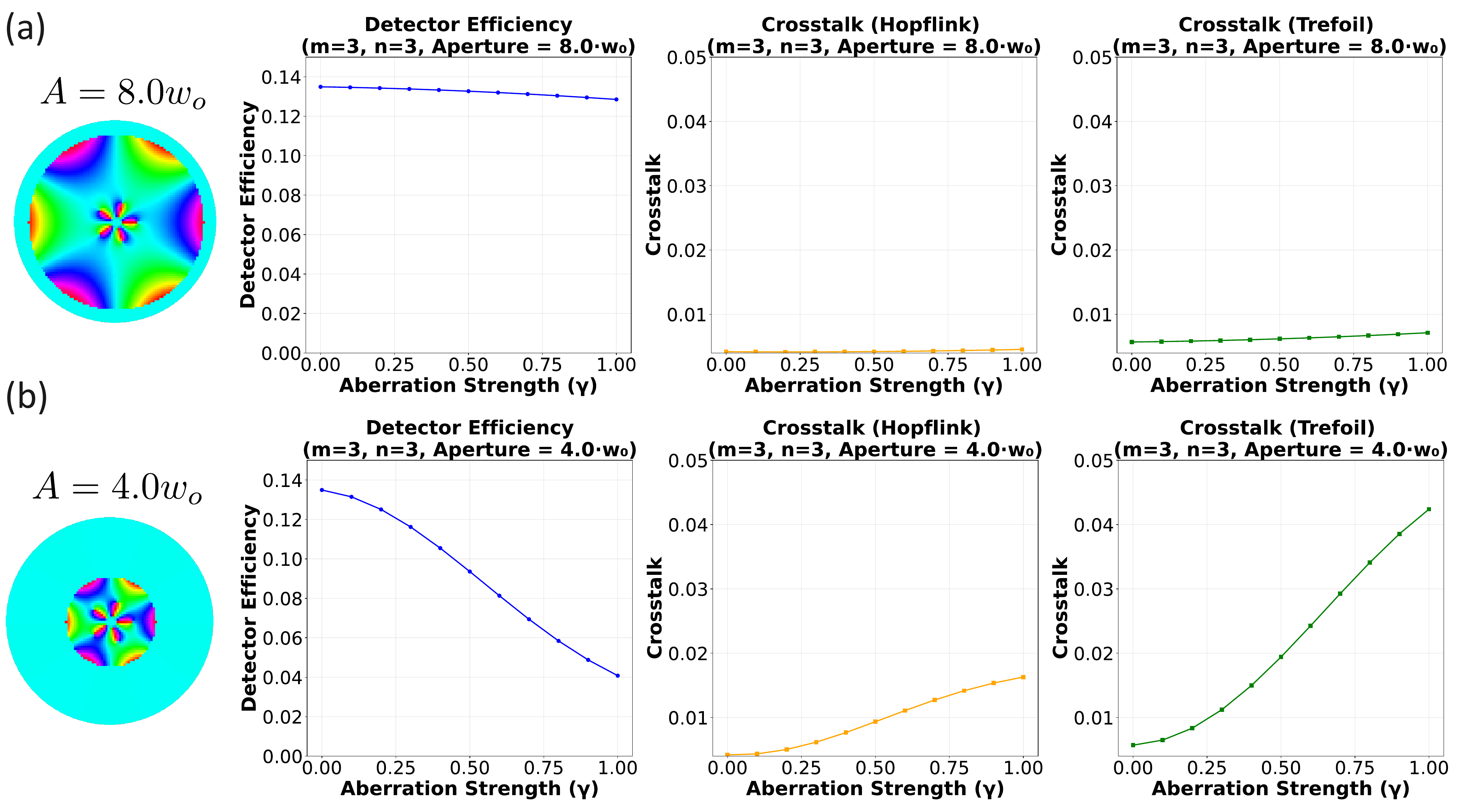}
    \caption{\textbf{Sorting aberrated knots.} We illustrate the impact of introducing trefoil ($Z_{3}^{3}$) aberration. The corresponding insets in (a) and (b) show the applied aberration with $\gamma=1.0$ and at different aperture sizes. Halving the aperture radius results in an enhanced perturbation, as the phase modulations more strongly affect the knotted singularities.  }
    \label{fig:sort_aberrated}
\end{figure*}
 
We show the results for different inputs in Figure~\ref{fig:sort_translation}, which, as expected, reveals a general decline in performance as we move farther from the center of the field. Curiously, however, the change in performance is not radially symmetric, and its allocation can vary between different knotted inputs. Interestingly, the locations of maximum detection efficiency and minimal crosstalk are located off-center from the detector.  

\textit{(3) Phase aberrations:} Lastly, we investigate the performance of the knot sorter under the influence of relative phase aberrations, encoded as an additional phase factor $\delta(\rho, \phi)$, induced by environmental disturbances or experimental imperfections and applied directly to the input knotted field. In remembrance of ~\cite{ak_paper}, we investigate the effect of individual Zernike modes, into which $\delta(\rho, \phi)$ can be decomposed. Specifically, the input knotted field is altered as $\tilde{\psi}_{\text{knot}}(\rho, \phi)  
= e^{i \pi \gamma Z_n^m}\psi(\rho, \phi),$ where
\begin{align}
Z_{n}^{m}(\rho/A, \phi) = 
\begin{cases}
    R^{m}_{n}(\rho/A)\cos{m\phi} & \text{if n}  > 0 \\
    R^{m}_{n}(\rho/A)\sin{m\phi} & \text{if n}  \leq 0
\end{cases},
\end{align}
and $R_{n}^{m}(\rho/A)$ refers to the radial polynomial defined in $0 \leq \rho \leq A$, $A$ refers to the aberration aperture, and $\gamma$ controls the aberration strength. In ~\cite{ak_paper}, it was shown that certain aberration effects can drastically change the knotted topology depending on the positioning of nontrivial wavefront distortions. Here, we also investigate modifications to the aperture and the strength of the aberration.

Figure~\ref{fig:sort_aberrated} illustrates how the sorter's performance changes under the influence of the trefoil ($Z^{3}_{3}$) aberration on a cinquefoil input. As the wavefront distortions lie about the circumference of the aberration mask, the knotted input remains largely invariant under the transformation with too large an aperture size, $A=8.0\,w_{o}$. Decreasing the aperture size by half causes both efficiency and sorting ability to drop dramatically. Similar trends in performance could be observed with Coma ($Z_3^{1}$), Secondary Astigmatism ($Z_4^{2}$), and Quadrafoil ($Z_4^{4}$) aberrations. With the former two, aberrations lie closer to the centre, so changing the aperture size does not dramatically affect trends in the sorting performance. That said, we observe that the knot's sorting ability decays with increasing aberration strength and decreasing aperture size. 

\textit{Real-world conditions.} We also benchmark the sorter's performance over a real-world channel. We investigate the underwater channel in~\cite{bouchard_quantum_2018} by constructing the experimentally retrieved phase modulation using the average Zernike coefficients and with an aperture size of $A=2.0w_{o}$. We evaluate the average sorting performance across all possible inputs and investigate the optimized mode sorters for $ D=2,3$. For $D=2$, we report average efficiencies and crosstalk of $26.46\%$ and $0.74 \%$ between the trefoil and cinquefoil, $23.93 \%$ and $1.21 \%$ between the trefoil and Hopf link, and $32.59\%$ between the Hopf link and cinquefoil. Among all of the $D=2$ sorters, we report average efficiency and crosstalk of $27.65 \%$ and $0.86 \%$, respectively. For $D=3$, we report an average efficiency of $13.8\%$ and a crosstalk of $0.48\%$. That the sorters' performance does not change is expected, given the low overall value of the Zernike coefficients. For both dimensions, the errors fall below the error threshold, attesting to the sorter's viability for implementing a secure communications link involving knots.

\section*{Conclusion}
We have demonstrated a scheme to sort a high-dimensional Alphabet of optical knots through a minimal number of phase elements. We show that we can highly distinguish up to three optical knotted fields. The crosstalk is sufficiently low to enable secure communication in high-dimensional quantum systems. Furthermore, we benchmark our knot sorter against various experimental imperfections and analyzed the evolution of sorting performance as a function of the type and severity of the imperfection. In particular, we investigated sorting performance in a real-world channel and found that the sorter maintains robust distinguishability. 

We provide an empirical prescription on knot shaping to achieve optimal theoretical sorting distinguishability. We stress, however, that choosing shape parameters that yield a closed optical knot is not entirely clear and could merit further investigation in the literature. We also note that we rely solely on information about the optical field at a particular plane of the knot, rather than on the knot's topology. More powerful would be to leverage information about the topological invariants of the optical knot for sorting~\cite{ferrer-garcia_polychromatic_2021, bar2025fast}. Future improvements of the knot sorter may be achieved through the introduction of an additional auxiliary output channel~\cite{goel2023simultaneously}. Rather than forcing partially overlapping knot states into one of the predefined output ports, the auxiliary channel can capture inconclusive contributions, thereby reducing classification errors and enhancing the overall performance of the sorting protocol. Moreover, one may consider using the Noll phase screen method, which represents arbitrary phase superpositions as a weighted sum of Zernike modes~\cite{peters2025structured}. Finally, the literature on linear diffractive neural networks has matured substantially, so there would be merit towards investigating optical knots implemented using these platforms~\cite{goel2023simultaneously,a_rocha_self-configuring_2025}.  

\section*{Code Avaliability}
The code used to produce the optimized phase holograms and to perform the simulations on the propagation of optical knotted fields in free space and within the sorter can be found at Ref. ~\cite{Jaouni_KnotSorter}.

\section*{Acknowledgments}
The authors acknowledge discussions with Drs Dilip Paneru and Manuel Ferrer, and Prof. Mehul Malik. This research is funded by the Canada Research Chairs program.

\bibliography{main}
\clearpage
\section*{Appendix}

\textit{Genetic Algorithm details.} We apply the GA in two steps to determine the optimized phase masks.  In the first $10^{4}$ iterations, we optimize the starting population with respect to the sorting performance, $F = \alpha \min{B_{n}} + (1-\alpha)\bar{B}$. Afterwards, we place additional emphasis on crosstalk and mode distinguishability by evaluating the full fitness metric as defined in Eq.~\ref{eq:optim_obj}. At each generation, we rank the phase masks in the population by their corresponding fitness metric and select the best to serve as 'parents' for the next generation using an exponential rank selection rule. We use an adaptive mutation rule in which chromosomes (here, the pixel-wise values of the phase screens) are mutated with a probability that is a function of the solutions' fitness values. Optimization continues for a specified number of generations, or until the fitness value converges after a minimum number of generations has elapsed. 
We provide a full list of parameters in Table~\ref{tab:ga_params}.

\textit{Knot equations.} We provide the explicit form of the nonzero polynomial weights, $c_{l}(\rho, a, b)$ in Table~\ref{tab:azimuthal_coefficients}

\begin{figure*}
\centering
\includegraphics[width=1.0\linewidth]{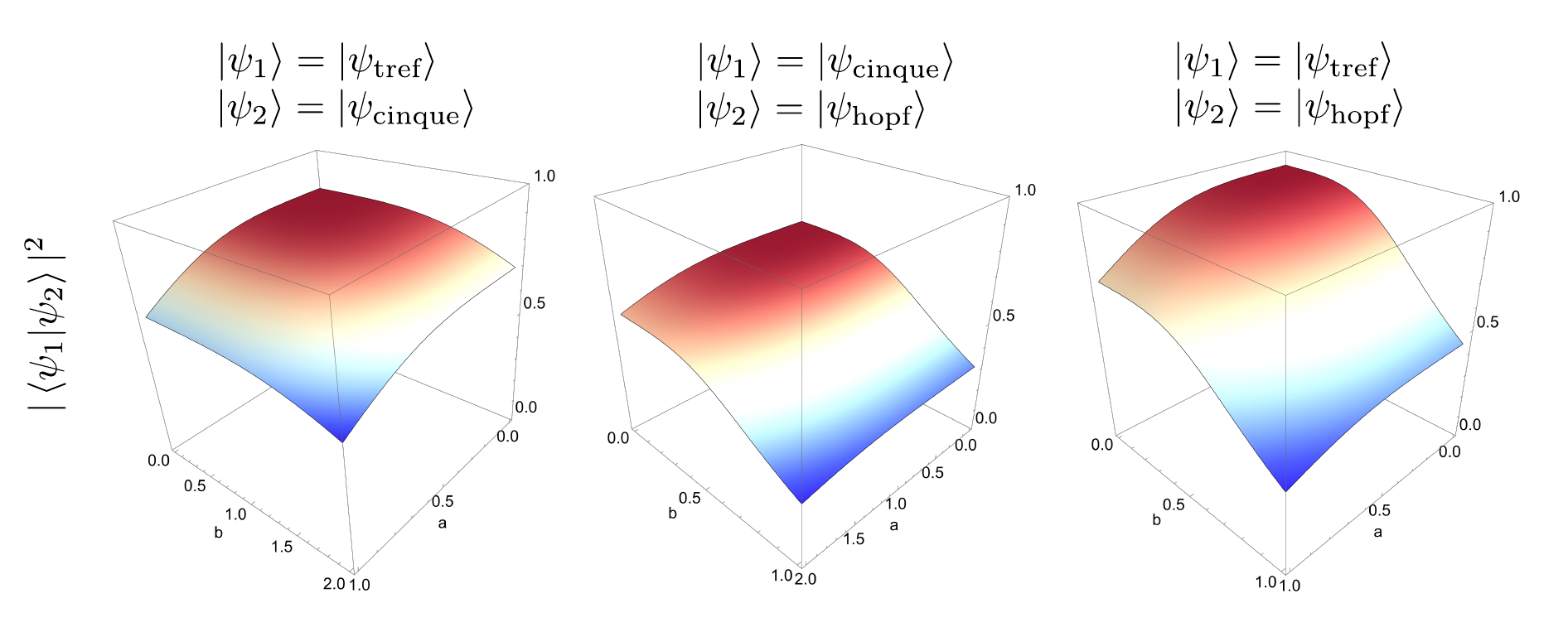}
    \caption{\textbf{Plot of the overlap between optical knots
with different shaping parameters.} The shape parameters are defined such that
$a_1 = b_1 = a$ for $\psi_1$, and $a_2 = b_2 = b$ for $\psi_2$. For both $\psi_1$ and
$\psi_2$, we have $s = 1.0$.}
    \label{fig:overlap_plots}
\end{figure*}

\textit{Shape parameters.} We plot the inner product between optical knots in Figure ~\ref{fig:overlap_plots} as a function of the shape parameters of each knot. The shape parameters are positive, real-valued numbers in principle, with higher-valued shape parameters yielding infinitesimally small inner product magnitudes. However, optical knots break apart beyond a certain regime of shapes.  We confirm this empirically by numerically propagating the optical knot under specific shape-parameter configurations and observing whether or not the singular skeleton closes.
\newpage
\begin{table}[!ht]
\centering
\renewcommand{\arraystretch}{1.0}
\begin{tabular}{l c}
\hline\hline
GA Parameter & Value \\
\hline
Number of Generations & $2.0\times10^{5}$ \\
Population Size & $20$ \\
Number of Parents & $4$ \\
Mutation Percentage & $15\%$ \\
High-Threshold Mutation Probability & $1\%$ \\
Low-Threshold Mutation Probability & $0.05\%$ \\
Minimum Mutation Value & $-\pi$ \\
Maximum Mutation Value & $+\pi$ \\
Crossover Type & Single-Point \\
\hline\hline
\end{tabular}
\caption{Parameters used for the GA optimization of the phase elements.  Our genetic algorithm was implemented in Python using the PyGAD package ~\cite{gad2024pygad}. All genetic instances were run on an AMD EPYC 9654 using the \textit{Narval} supercluster.}
\label{tab:ga_params}
\end{table}
\begin{table}[!h]
\centering
\renewcommand{\arraystretch}{1.0}
\begin{tabular}{c c l}
\hline\hline
Type & $\ell$ & $c_{\ell}(\rho, a, b)$ \\
\hline
\multirow{3}{*}{Hopf link}
  & $0$  & $1-2\left(1+a^{2}-b^{2}\right)\rho^{2}+\rho^{4}$ \\
  & $+2$ & $-\rho^{2}(a+b)^{2}$ \\
  & $-2$ & $-\rho^{2}(a-b)^{2}$ \\
\hline
\multirow{3}{*}{Trefoil}
  & $0$  & $\rho^{6}-\rho^{4}-\rho^{2}+1-4\rho^{3}\left(a^{2}-b^{2}\right)$ \\
  & $+3$ & $-2\rho^{3}(a+b)^{2}$ \\
  & $-3$ & $-2\rho^{3}(a-b)^{2}$ \\
\hline
\multirow{3}{*}{Cinquefoil}
  & $0$  & $1+\rho^{2}-2\rho^{4}-2\rho^{6}+\rho^{8}+\rho^{10}-16\rho^{5}\left(a^{2}-b^{2}\right)$ \\
  & $+5$ & $-8\rho^{5}(a+b)^{2}$ \\
  & $-5$ & $-8\rho^{5}(a-b)^{2}$ \\
\hline\hline
\end{tabular}
\caption{Explicit coefficients for the polynomial specifying each optical knot.}
\label{tab:azimuthal_coefficients}
\end{table}

\end{document}